\documentclass[10pt,letterpaper,twocolumn]{article} 

\usepackage{ol2}
\usepackage[draft]{hyperref}
\usepackage{amsmath}

\begin{document}

\twocolumn[ 

\title{Quantum interference and Exceptional Points}


\author{Stefano Longhi}

\address{Dipartimento di Fisica, Politecnico di Milano and Istituto di Fotonica e Nanotecnologie del Consiglio Nazionale delle Ricerche, Piazza L. da Vinci 32, I-20133 Milano, Italy (stefano.longhi@polimi.it)}

\begin{abstract}
Exceptional points (EPs), i.e. branch point singularities of non-Hermitian Hamiltonians, are ubiquitous in optics. So far, the signatures of EPs have been mostly studied assuming classical light.
 In the passive parity-time ($\mathcal{PT}$) optical coupler, a fingerprint of EPs resulting from the coalescence of two resonance modes is a qualitative change of the photon decay law, from damped Rabi-like oscillations to transparency, as the EP is crossed by increasing the loss rate. However, when probed by non-classical states of light, quantum interference can hide EPs. Here it is shown that, under excitation with polarization-entangled two-photon states, EP phase transition is smoothed until to disappear as the effective particle statistics is changed from bosonic to fermionic.
 
\end{abstract}

 ] 

Recently, optical systems operating close to an exceptional point (EP) are attracting a great interest with a wealth of applications in photonics. Among them one should mention unidirectional transparency \cite{r1,r2,r3}, laser mode control \cite{r4,r5,r6}, light structuring \cite{r7}, parity-time  ($\mathcal{PT}$) laser-absorber devices \cite{r8,r9}, optical sensing \cite{r10,r11,r12,r13}, light stopping \cite{r14}, etc.. EPs, also referred to as non-Hermitian degeneracies, are singular points in parameter space of a non-Hermitian Hamiltonian where two (or more) eigenvalues and their corresponding eigenstates coalesce under a system parameter variation \cite{r16,r17}. In optics, a paradigmatic system exhibiting an EP is the $\mathcal{PT}$-symmetric optical coupler, i.e. a system of two coupled waveguides with balanced optical gain and loss \cite{r18}. In the so-called passive $\mathcal{PT}$ optical coupler \cite{r19},  EP corresponds to the coalescence of the two leaky supermodes of the coupler when the unbalanced loss rates in the two waveguides equals the coupling rate. Like for similar non-Hermitian resonance crossing occurring in open quantum systems \cite{r20}, the EP separates two qualitatively distinct regimes of photon decay \cite{r20,r21,r22}, from damped Rabi-like oscillations to loss-induced transparency, which have been experimentally observed in microwave and optical experiments \cite{r19,r21}. Other fingerprints of EPs are the enhanced sensitivity to variation of parameters \cite{r10,r11,r12,r13,r23} and chirality observed when an EP is dynamically encircled \cite{r24,r25}. The signatures of EPs have been mostly explored assuming classical light. A few works have considered system behavior near an EP in second-quantization framework \cite{r26,r27,r28,r29,r30,r30bis}, focusing on the impact of EPs on quantum noise, especially in the presence of some gain (population inversion) in the system. A fundamental property of linear passive optical systems probed by nonclassical states of light is to show multi-photon quantum interference \cite{r31}. So far, the interplay between quantum interference and EPs has been overlooked.\\
{In this Letter we consider quantum interference effects in an optical system near an EP, namely a passive $\mathcal{PT}$ optical waveguide coupler, and show that, when the coupler is excited by a polarization-entangled two-photon state,  the phase transition at EP crossing, measured by the two-photon survival probability, undergoes a qualitative change and becomes hidden as the particle statistics is tuned from bosonic to fermionic \cite{r32,r33,r34}.}\\ 
\begin{figure}[htb]
\centerline{\includegraphics[width=8.4cm]{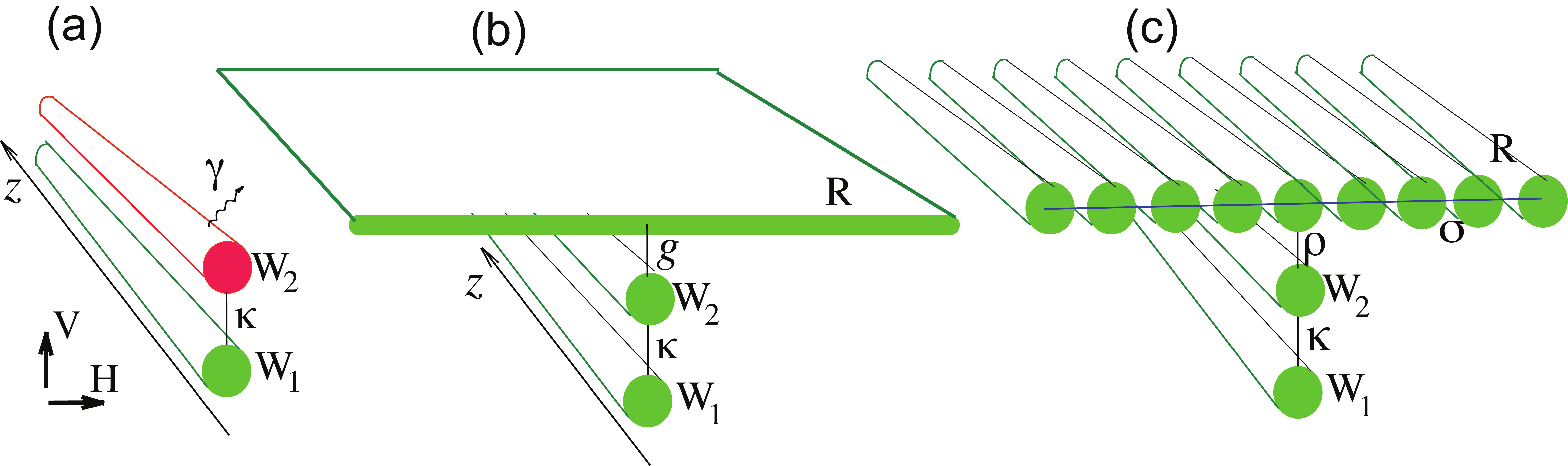}} \caption{ \small
(Color online) (a) Schematic of a passive $\mathcal{PT}$ optical directional coupler. Optical losses in waveguide $\rm{W}_2$ can arise from material absorption [as in (a)], or from evanescent coupling to a highly-multimode structure R, such as a slab waveguide [panel (b)] or a waveguide lattice [panel (c)].}
\end{figure} 
The passive $\mathcal{PT}$ optical coupler \cite{r19} consists of two coupled optical waveguides $\rm{W}_1$ and $\rm{W}_2$, the former being lossless while the latter exhibiting some linear optical loss due to material absorption [Fig.1(a)] or to evanescent mode coupling with surrounding scattering modes [Figs.1(b) and (c)]. We assume that each of the two waveguides sustains two linearly-polarized nearly-degenerate modes, i.e. we neglect waveguide birefringence. For excitation with classical light, using coupled-mode theory  light propagation in the coupler is described by the following equations for the mode amplitudes $c_1$ and $c_2$ in either one of the two polarization state 
\begin{equation}
i \frac{d}{dz}
\left(
\begin{array}{c}
c_1 \\
c_2
\end{array}
\right)=
\left( 
\begin{array}{cc}
\beta_1 & \kappa \\
\kappa & \beta_2-i \gamma
\end{array}
\right)  \left(
\begin{array}{c}
c_1 \\
c_2
\end{array}
\right) \equiv M \left(
\begin{array}{c}
c_1 \\
c_2
\end{array}
\right).
\end{equation}
In Eq.(1), $\beta_1$, $\beta_2$ are the propagation constants of the modes in the two waveguides, $\gamma$ is the loss rate of the lossy waveguide, and $\kappa$ is the coupling constant. Assuming $\beta_1=\beta_2 \equiv \beta$, the coupler sustains two leaky supermodes with propagation constants given by
\begin{equation}
\lambda_{1,2}=\beta-i \frac{\gamma}{2} \pm \sqrt{\kappa^2-(\gamma/2)^2}.
\end{equation} 
For a small loss $\gamma < 2 \kappa$, the two supermodes have the same damping rate $\gamma/2$ but different phase velocities, resulting in a damped oscillatory (Rabi-like) decay of the optical power $P(z)=|c_1(z)|^2+|c_2(z)|^2$ along the coupler \cite{r21}.  At $\gamma= 2 \kappa$, an EP arises, with the coalescence of the two eigenvalues $\lambda_+=\lambda_-=\beta-i \gamma/2$ and corresponding eigenvectors of the matrix $M$. At the EP, the anomalous exponential-power decay law $P(z) \sim z^2 \exp(-\gamma z)$ is observed \cite{r20,r22}. Above the EP, i.e. for $\gamma> 2 \kappa$, the two supermodes have the same phase velocity but different damping rates. In particular, in the strong loss limit $\gamma \gg \kappa$ one of the two supermodes has a vanishing decay rate due to loss-induced decoupling of the two waveguides, an effect referred to as loss-induced transparency \cite{r19}. For non-birefringent waveguides, the results are independent of the polarization state excitation, i.e. the electric field polarized along the horizontal (H) or vertical (V) transverse axes. {For a rather arbitrary input excitation of the coupler, both supermodes are excited and their superposition clearly yields a deviation from a pure exponential decay law, with a qualitative change of the decay behavior as the EP is crossed by increasing the loss rate $\gamma$.} Some examples  of optical power decay behaviors below, at and above the EP are shown in Fig.2, corresponding to  two different excitation conditions of the coupler.
\par
\begin{figure}[htb]
\centerline{\includegraphics[width=8.4cm]{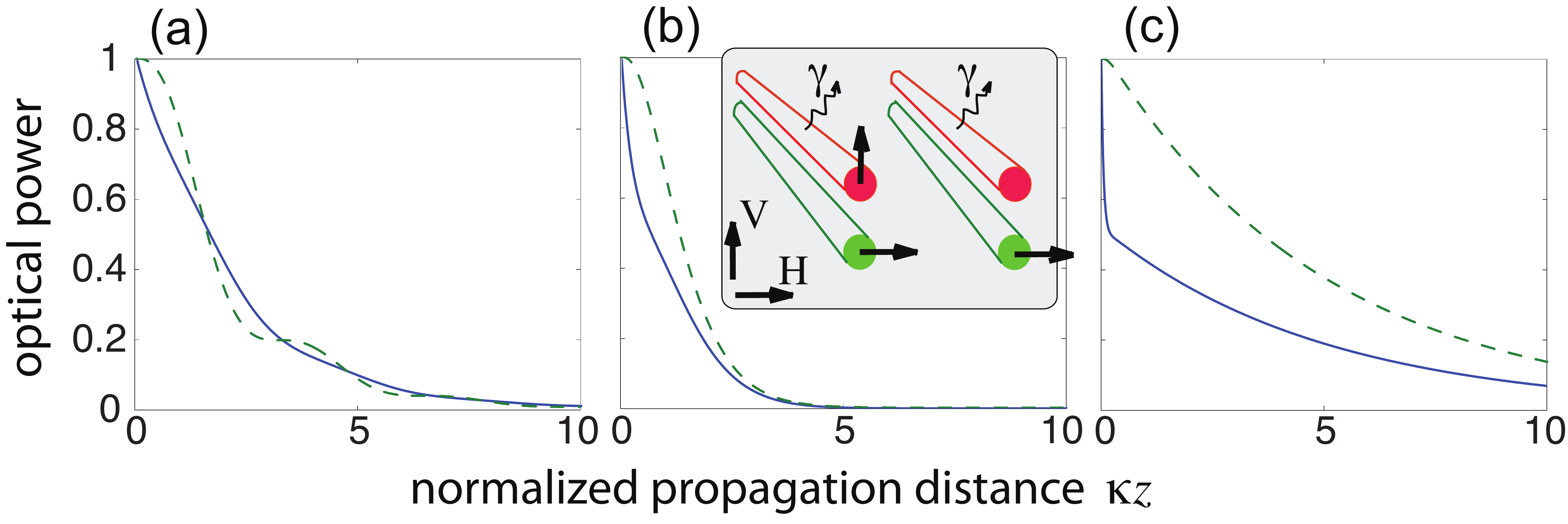}} \caption{ \small
(Color online) Behavior of the total optical power $P(z)$, normalized to its input value, versus the propagation distance $z$ in a passive $\mathcal{PT}$ optical coupler  for (a)  $\gamma / \kappa=0.5$, (b) $\gamma / \kappa=2$ (at the EP), and (c) $\gamma / \kappa=10$. Solid curves refer to balanced input excitation with orthogonal polarization modes {(incoherent-like excitation)} $c_1^{(H)}(0)=c_2^{(V)}(0)=1/ \sqrt{2}$, $c_1^{(V)}(0)=c_2^{(H)}(0)=0$, whereas dashed curves refer to single waveguide excitation $c_1^{(H)}(0)=1$, $c_2^{(H,V)}(0)=c_1^{(V)}(0)=0$ (left and right pictures in the inset).}
\end{figure} 
To study propagation of non-classical light in the passive $\mathcal{PT}$ optical coupler, one should consider the quantum version of Eq.(1), i.e. two-coupled quantum oscillators, one of which interacting with a reservoir of bosonic oscillators that gives rise to photon loss and decoherence. Standard procedures in quantum optics are based on the master equation for the reduced density operator \cite{r35} or the use of Langevin equations \cite{r29,r36}, after the reservoir degrees of freedom are eliminated in the Born-Markov approximation. To study quantum interference effects,  it is more convenient to consider photon state propagation in the entire waveguide-reservoir system using the method described in \cite{r37}, in which the optical losses are emulated by evanescent coupling of $\rm{W}_2$  to a a set of scattering modes of a highly-multimode structure R, such as an optical slab or a waveguide lattice \cite{r38}; Fig.1(b) and (c). The second-quantization Hamiltonian for the photon field reads $\hat{H}=\hat{H}_W+\hat{H}_R+\hat{H}_{coup}$, where
\begin{eqnarray}
\hat{H}_W & = & \hbar c \left( \beta_1 \hat{c}_1^{\dag}\hat{c}_1+  \beta_2 \hat{c}_2^{\dag}\hat{c}_2 \right) +\hbar c \kappa ( \hat{c}_1^{\dag} \hat{c}_2+ \hat{c}_2^{\dag} \hat{c}_1)\\
\hat{H}_R & = & \hbar c \int dk \beta(k) \hat{a}^{\dag}(k) \hat{a}(k)  \\
\hat{H}_{coup} & = & \hbar c \int dk \left( g(k) \hat{c}_2^{\dag} \hat{a}(k) + g^*(k) \hat{c}_2 \hat{a}^{\dag}(k)  \right)  
\end{eqnarray}
describe the quantum field in the lossless optical coupler ($\hat{H}_W$), in the multimode waveguide R that acts as the reservoir ($\hat{H}_R$), and their coupling ($\hat{H}_{coup}$). In the above equations, $\hat c_{1}$ and $\hat c_2$ are the destruction operators for individual modes in the two waveguides ${\rm W}_1$ and  ${\rm W}_2$ in a given polarization state (either H or V), $\hat{a}(k)$ are the destruction operators of the scattering modes in the structure R, defined by some continuous index $k$, $\beta=\beta(k)$ is the dispersion relation of modes in R, $g(k)$ is the coupling constant between the mode of waveguide W$_2$ and the scattering mode in R with index $k$, and $c$ is the speed of light. Taking into account that $z=ct$, the Heisenberg equations of motion for the destruction operators read
\begin{eqnarray}
i \frac{d \hat{c}_1}{dz}  & = &  \beta_1 \hat{c}_1+\kappa \hat{c}_2 \\
i \frac{d \hat{c}_2}{dz}  & = &  \kappa \hat{c}_1+\beta_2 \hat{c_2}+\int dk g(k) \hat{a}(k,z) \\
i \frac{d \hat{a}}{dz}  & = &  \beta(k) \hat{a}(k,z)+g^*(k) \hat{c}_2.
\end{eqnarray}
{Since Eqs. (6-8) are linear, their solution is of the form}
\begin{equation}
\hat{c}_1(z)  = S_{1,1}(z) \hat{c}_1(0)+S_{1,2} \hat{c}_2(0)+ \int dk S_1 (k,z) \hat{a}(k,0) \;\;
\end{equation}
\begin{equation}
\hat{c}_2(z)  =  S_{2,1}(z) \hat{c}_1(0)+S_{2,2}(z) \hat{c}_2(0)+\int dk S_2(k,z) \hat{a}(k,0) \;\;
\end{equation}
where the scattering amplitudes $S_{n,l}(z)$ ($n,l=1,2)$, $\mathcal{S}_1(k,z)$ and $\mathcal{S}_2(k,z)$ can be computed from the corresponding classical problem, i.e. when the operators in Eqs.(6-8) are considered as $c$-numbers. In the weak-coupling (markovian) approximation and for an unstructured continuum of modes in R, the elements $S_{l,n}(z)$ ($n,l=1,2)$ can be calculated 
{following a standard procedure: after elimination of the  $c$-number amplitudes ${a}(k,z)$ from the dynamics via a formal integration of Eq.(8), coupled integro-differential equations for ${c}_{1,2}$ are obtained, which reduce to the ordinary differential equation (1) in the markovian (short memory time) approximation (see e.g. 
 \cite{r37,r37bis} for technical details). The scattering amplitudes $S_{n,l}(z)$ are then given by}
\begin{equation}
\left(
\begin{array}{cc}
S_{1,1} & S_{1,2} \\
S_{2,1} & S_{2,2}
\end{array}
\right)
\simeq \exp(-i M z)
\end{equation}
\begin{figure}[htb]
\centerline{\includegraphics[width=8.4cm]{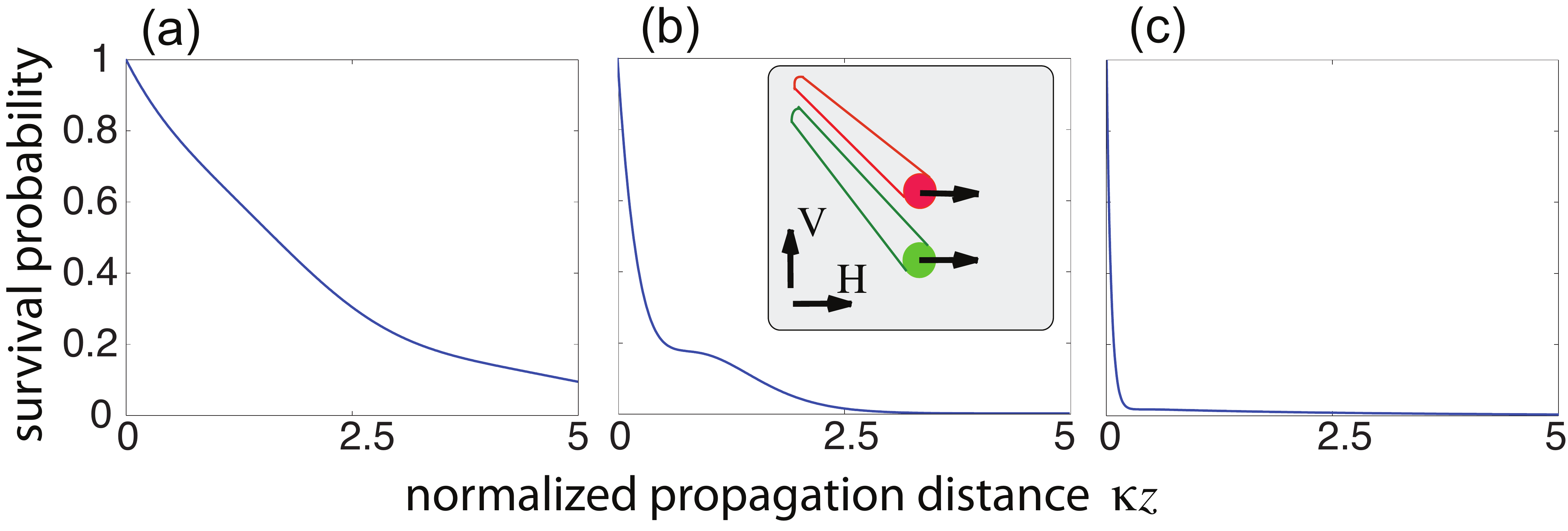}} \caption{ \small
(Color online) Behavior of the survival probability $P(z)$ versus propagation distance $z$ in a passive $\mathcal{PT}$ optical coupler excited by the two-photon state (16)  for (a)  $\gamma / \kappa=0.5$, (b) $\gamma / \kappa=2$, and (c) $\gamma / \kappa=10$.}
\end{figure}
where $M$ is defined in Eq.(1) and where  \cite{r37,r37bis}
\begin{equation}
\gamma= \int_0^{\infty} d \tau \int dk |g(k)|^2 \exp[i \beta_2 \tau -\beta(k)\tau ] .
\end{equation}
The real and imaginary parts of $\gamma$ define the loss rate and propagation constant shift (Lamb shift) of waveguide W$_2$ due to the coupling with the continuum. In the following, for the sake of simplicity we will assume a vanishing Lamb shift, i.e. $\gamma$ real.
For example, if R is a waveguide lattice with a tight-binding band of with $4 \sigma$ [Fig.1(c)] and the waveguides in the lattice have the same propagation constant $\beta_2$ than the waveguide W$_2$, one has $\beta(k)= \beta_2+2 \sigma \cos(k)$, $g(k)=\rho / \sqrt{2 \pi}$ ($ -\pi \leq k < \pi$), and from Eq.(12) one obtains \cite{r37bis}
\begin{equation}
\gamma=\rho^2 /(2 \sigma ).
\end{equation}
where $\rho$ is the coupling between waveguide W$_2$ and the array [see Fig.1(c)]. 
Let us indicate by $|\psi(z=0) \rangle$ the vector state of the quantum field at the excitation pane $z=0$ of the coupler. The state $|\psi(z=0) \rangle$ is defined by a superposition of photon number states $(1/\sqrt{n!m!}) \hat{c}_1^{\dag n} \hat{c}_1^{\dag m} | 0 \rangle$. In the Schr\"odinger picture, the vector state of the quantum field $|\psi(z) \rangle$  at the propagation distance $z$ is obtained from the expression of  $|\psi(z=0) \rangle$ by the formal substitutions \cite{r37}
\begin{eqnarray}
\hat{c}^{\dag}_1 & \rightarrow & S_{1,1} (z) \hat{c}_1^{\dag}+ S_{2,1} (z) \hat{c}_2^{\dag}+... \\
\hat{c}^{\dag}_2 & \rightarrow & S_{1,2} (z) \hat{c}_1^{\dag}+ S_{2,2} (z) \hat{c}_2^{\dag} +...
\end{eqnarray}
where the dots involve operators $\hat{a}^{\dag}(k)$ of continuum. {When the optical coupler is excited by nonclassical light, for example by photon number states, rather generally the evolution along $z$ of the  mean photon number $\langle  n \rangle=\langle  \hat{c}_1 \hat{c}^{\dag}_1+ \hat{c}_2 \hat{c}^{\dag}_2 \rangle $ in the two waveguides follows the classical decay law, and therefore the transition at the EP can be observed like for classical light (coherent state) excitation. To highlight the interplay of EP and quantum light, we need to look at the photon statistics in the two waveguides and how it changes as the EP is crossed. In particular, for excitation with a quantum state with defined number of photons an observable quantity is provided by the non-decaying (survival) probability $P(z)$, defined as the probability that at the propagation distance $z$ none of the initially injected photons in the coupler have decayed into the continuum. Let us specifically focus our attention to the excitation of the optical coupler with a two-photon state.} As a first example, let us assume that the coupler is excited by 
{two indistinguishable photons, one photon injected in each waveguide with the same polarization state (either H or V)}, i.e. let us assume
\begin{equation}
| \psi(z=0) \rangle =\hat{c}_1^{\dag} \hat{c}_2^{\dag} | 0 \rangle.
\end{equation}
\begin{figure}[htb]
\centerline{\includegraphics[width=8.4cm]{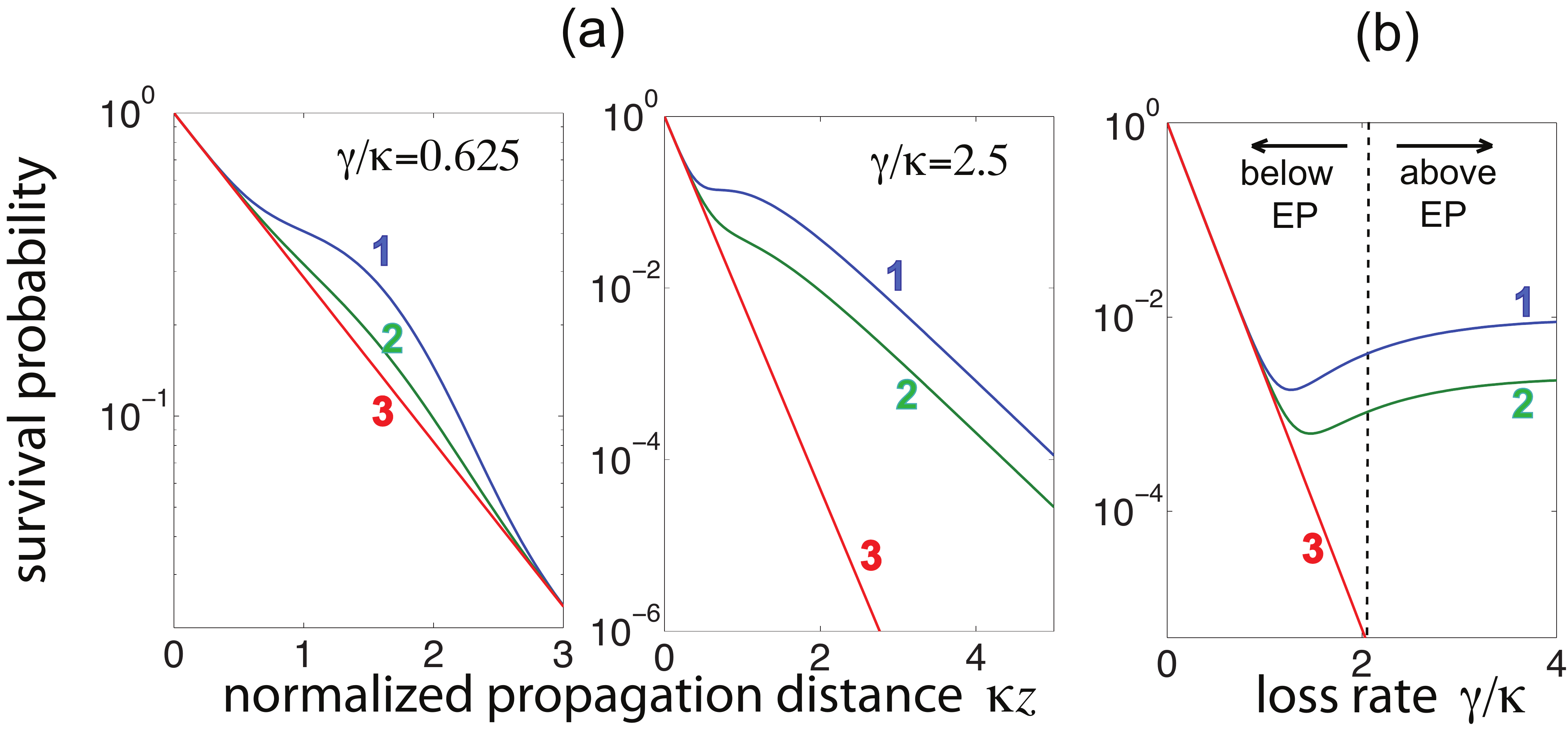}} \caption{ \small
{(Color online) (a) Behavior of the survival probability $P(z)$ versus $z$ on a log scale in a passive $\mathcal{PT}$ optical coupler excited by the polarization-entangled two-photon state (19) below ($\gamma/ \kappa=0.625$) and above ($\gamma/ \kappa=2.5$) the EP and for a few values of the phase $\varphi$ (curve 1: $\varphi=0$; curve 2: $\varphi=2 \pi/3$; curve 3: $\varphi=\pi$). (b) Behavior of the survival probability $P$, at fixed position $z_0=3 \kappa$, versus loss rate $\gamma$.}}
\end{figure}
According to the substitutions (14) and (15), the vector state at plane $z$ is given by
\begin{eqnarray}
| \psi(z) \rangle & = & S_{1,1}S_{1,2} \hat{c}_1^{\dag 2} | 0 \rangle + S_{2,1}S_{2,2} \hat{c}_2^{\dag 2} | 0 \rangle \\
 &  + & \left\{ S_{1,1}S_{2,2}+S_{2,1} S_{1,2} \right\} \hat{c}_1^{\dag} c_2^{\dag} | 0 \rangle  +... \nonumber
\end{eqnarray}
where the dots denote terms involving $\hat{a}^{\dag}(k)$ operators. {For such an excitation, it can be readily shown that the mean number of photons $\langle n \rangle$ that remains in the coupler follows the classical decay behavior shown by the solid curves in Fig.2 (balanced incoherent-like excitation). On the other hand the survival probability is given by}
\begin{eqnarray}
P(z)& = & 2|S_{1,1}(z)S_{1,2}(z)|^2+2|S_{2,1}(z)S_{2,2}(z)|^2+ \nonumber \\
& + & |S_{1,1}(z)S_{22}(z)+S_{1,2} (z)S_{2,1}(z)|^2.
\end{eqnarray}
A typical behavior of $P(z)$, for a loss rate $\gamma$ below, at and above the EP, is shown in Fig.3. 
Like for the mean photon number $\langle n \rangle$, the decay of $P(z)$ is not exponential and a qualitative change  is observed when crossing the EP. {Comparing Figs.2 and 3, the main difference is that the survival probability,  for large loss rates $\gamma$ [panels (c)], rapidly drops to a much smaller value than the mean photon number. This is simply due to the fact that $P(z)$ requires {\em both} photons to remain in the coupler: for a large loss rate this is a less probable event because in state (16) the lossy waveguide is excited. On the other hand, for large losses one photon is likely to stay in the non-lossy waveguide, which is effectively decoupled from the lossy one: hence $\langle n \rangle$ [Fig.2(c), solid curve] does not show the abrupt drop of $P(z)$ [Fig.3(c)].\\
 A much more interesting result is obtained when the coupler is excited by a polarization-entangled two-photon state}
\begin{equation}
| \psi(z=0) \rangle= \frac{1}{\sqrt{2}} \left( \hat{c}_1^{(H) \dag} \hat{c}_2^{(V) \dag} + \exp(i \varphi) \hat{c}_1^{(V) \dag} \hat{c}_2^{(H) \dag}\right) |0 \rangle
\end{equation}
where $\varphi$ is a phase shift and $\hat{c}_{1,2}^{(H) \dag}$, $\hat{c}_{1,2}^{(V) \dag}$ are the creation operators of photon waveguide modes with H and V polarization,  respectively. 
By changing the phase $\varphi$, from $\varphi=0$ to $\varphi= \pi$, the polarization-entangled two-photon state (19) effectively emulates quantum interference of bosonic and fermionic particles, with anyonic statistics for intormediate values \cite{r32,r33}. For polarization-independent scattering amplitudes, the survival probability reads
\begin{eqnarray}
P(z) & = & 2 \cos^2 \left( \frac{\varphi}{2} \right) \left( |S_{1,1}S_{1,2}|^2+|S_{2,1}S_{2,2}|^2 \right) \nonumber \\
& + &  |S_{1,1}S_{2,2}|^2+|S_{1,2}S_{2,1}|^2 \\
& + & \cos \varphi \left(S_{1,1}S_{2,2}S_{1,2}^*S_{2,1}^*+ S_{1,1}^*S_{2,2}^*S_{1,2}S_{2,1} \right). \nonumber
\end{eqnarray}
{A typical behavior of $P(z)$ on a log scale, corresponding to a few values of the phase $\varphi$, is shown in Fig.4(a) for a loss rate below and above the EP. The behavior of the survival probability, for a fixed propagation distance $z=z_0$ versus the loss rate $\kappa / \gamma$, is shown in Fig.4(b). The main intriguing effect is found for $\varphi=\pi$: the decay of $P(z)$ is always exponential and no signature of EP crossing is found: while loss-induced transparency is observed for bosonic statistics, this is not the case for fermionic one [see Fig.4(b)]. This property can be readily proven by observing that, for $\varphi=\pi$, from Eq.(20) one obtains}
\begin{equation}
P(z)=|S_{1,1}S_{2,2}-S_{1,2}S_{2,1}|^2 \simeq \exp(- 2 \gamma z)
\end{equation}
where we used Eq.(11) and the property 
$ {\rm det} \left\{ \exp(-iMz) \right\}=\exp \{-iz  Tr(M) \}$. Equation (21) states that the decay law $P(z)$ is {\em always} an exponential one and it is insensitive to the  fact that the $\mathcal{PT}$ coupler is below or above the EP. {Another remarkable result is that, above the EP ($\gamma > \simeq 2 \kappa$), the  probability $P(z_0; \varphi)$ undergoes a large change, by more than two orders of magnitudes, as the phase $\varphi$ varies from $0$ to $\pi$ [see Fig.4(b)], which could hold major interest in sensing applications based on quantum light.} 
 \begin{figure}[htb]
\centerline{\includegraphics[width=8.4cm]{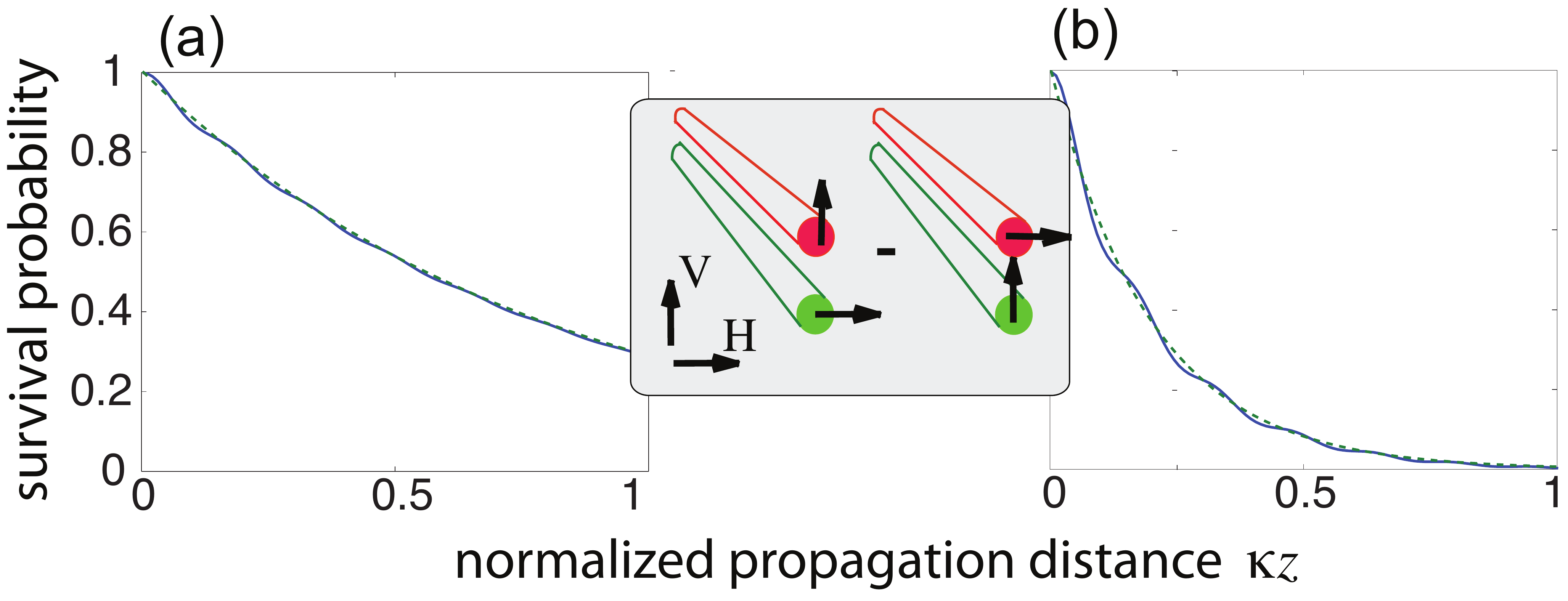}} \caption{ \small
(Color online) Behavior of the survival probability versus $z$ in a passive $\mathcal{PT}$ optical coupler excited by antisymmetric polarization-entangled two-photon state. The optical setup is shown in Fig.1(c), corresponding to an effective loss rate $\gamma \simeq \rho^2/(2 \sigma)$ in waveguide W$_2$. Parameter values are $\sigma / \kappa=20$ and (a) $\rho / \kappa=5$ (corresponding to $\gamma / \kappa \simeq 0.625$, below the EP), and (b) $\rho / \kappa=10$ (corresponding to $\gamma / \kappa \simeq 2.5$, above the EP). Dashed curves, almost overlapped with the solid ones, depict the exact exponential behavior of the survival probability in the markovian approximation [Eq.(21)].}
\end{figure}
 We note that such results are exact ones in the markovian approximation. For fermionic statistics, non-markovian effects, arising from a moderate coupling or for a structured continuum R, are expected to introduce some deviations of the decay law from an exponential one, however such deviations are largely insensitive to the EP phase transition.\\
The predicted effect could be experimentally demonstrated on an integrated quantum photonic platform \cite{r32,r34,r39} using the waveguide configuration of Fig.1(c) . In this setup, the passive optical coupler is side-coupled to a waveguide lattice that introduces effective loss for waveguide W$_2$, which can be tailored by varying the waveguide spacing. Figure 5 shows numerical results of the behavior of the survival probability $P(z)$, as obtained by exact numerical computation of the scattering amplitudes $S_{n,l}$ beyond the markovian approximation, for fermionic statistics and for two different coupling values $\rho$ corresponding to a ratio $\gamma / \kappa$ below [Fig.5(a)] and above [Fig.5(b)] the EP. The dashed curves in the plots, almost overlapped with the solid ones, show the exponential decay Eq.(21) predicted in the markovian approximation. The numerical results clearly indicate that, even beyond the markovian limit, EP crossing is hidden.\\
In conclusion, we studied EP phase transitions in passive photonic structures probed at the few photon level, and showed that quantum interference can smooth and even hide the phase transition arising from non-Hermitian degeneracy crossing.  The interplay between quantum interference and EP could be of potential interest for precision measurements and optical sensing applications, where an optical structure near an EP is probed using quantum light.

\end{document}